# Temperature Dependence of the Yield Strength of Aluminum Thin Films: Multiscale Modeling Approach

Kamyar Davoudi [*]

School of Engineering and Applied Sciences, Harvard University, Cambridge, MA 02138, USA

**Abstract**

Modeling deformation at elevated temperatures using discrete dislocation dynamics (DDD) is a recent area of high interest. However, the literature dedicated to this subject fails to address the variations of DDD parameters with temperature. This study aims to investigate the effect of temperature on the yield strength of aluminum thin films in two-dimensional DDD simulations. To this end, the temperature dependence of DDD parameters has been studied using molecular dynamics, three-dimensional DDD simulations, and the existing experimental results. Based on these calculations, we observed 18% decrease in the yield strength when temperature was increased from 100 K to 600 K.

Discrete dislocation dynamics (DDD) is a powerful tool that has been successfully applied to simulate plastic deformation in crystalline materials specially at the micron and submicron scales in various modeling problems. In this method, the material is modeled as a continuous medium containing dislocations as the main plastic carriers, where evolution of dislocations is tracked. In the formulation of the DDD approach, it is assumed that the motion of a dislocation of unit length is governed by:

$$m^* \dot{v} + Bv = F(t) \qquad (1)$$

where $v$ is the dislocation velocity, $m^*$ is the effective mass, $B$ is the drag coefficient, and $F(t)$ is the driving force arising from externally applied stresses, image stresses, self-stress, interactions with other defects, and the lattice friction (Peierls stress). In this method, long range interactions between dislocations are calculated through theory of elasticity; short range interactions are determined based on local rules.

DDD is implemented either in two- or three-dimensional frameworks. Three-dimensional DDD (3d DDD) includes dislocations in realistic geometries and many dislocation phenomena such as junction formation and cross-slip. 2d simulations include only infinitely long edge dislocations. In 2d approach, Frank-Read (FR) sources are represented by points; when the resolved shear stress exceeds a prescribed nucleation stress during a period of time, called nucleation time, the FR source emits a dislocation dipole. Therefore, some important mechanisms such as cross-slip are absent in 2d DDD, which limits the problems to which this approach can be applied. 3d DDD can capture physics of dislocation motion more accurately. However, 3d calculations are complex and computationally expensive [1]. Many of the current 3d DDD codes, such as ParaDis, originally developed in Lawrence Livermore National Lab, can model only single crystals. Therefore, compared to 2d DDD simulations, most 3d DDD studies are limited to smaller plastic strains, lower dislocation density, or often simpler geometries. 2d DDD can be coupled with other computationally intense simulation techniques, such as phase field method more easily. That is why 2d DDD is still vastly used for modeling plasticity.

---

[*] *Email address:* davoudi@seas.harvard.edu

In recent years, deformation at elevated temperatures and incorporating dislocation climb into DDD codes have come to receive much attention [2–14]. Increasing temperate affects several parameters of simulations including elastic constants, drag coefficients in both 2d and 3d DDD, and critical nucleation stress, $\tau_{nuc}$, and nucleation time of Frank-Read sources, $t_{nuc}$, in 2d DDD. Also the Burgers vector negligibly increases by 2% when the temperature increases from 0 K to 900 K [15]. The variation of the drag coefficient with temperature has only been shown in some of the above cited papers, and assumed to vary linearly with the temperature. This, as we will see later, is not an accurate assumption as linear relationships don't hold true in a long range of temperature changes. Other changes have not been addressed in these papers.

In this letter, we investigate how parameters of 2d simulations are affected by temperature elevation in aluminum. Because aluminum is a widely-used material in industry, is highly under the influence of temperature changes, and is most studied in the works assessing climb in DDD simulations, we consider this fcc material. We focus on early stages of plastic deformation where thermally activated mechanisms do not play key roles. To this end, we first consider how elastic constants change based on the existing experimental results. Second, we use molecular dynamics (MD) to calculate changes of the drag coefficients with temperature. Effects of temperature on critical nucleation stress and nucleation time are assessed using 3d DDD. At the end, to emphasize on the importance of variation of these parameters, we study the effects of temperature on the yield strength.

The variations in elastic constants of a single crystalline aluminum with temperature have been determined experimentally [16–19] and computationally [20]. The isotropic elastic moduli for a polycrystalline aggregate can be estimated using the Voigt or Reuss models. The shear modulus $\mu$ and Poisson's ratio $v$ of polycrystalline aluminum are calculated using the experimental data given by Kamm and Alers [17] and Gerlich and Fisher [18] (Fig. 1). The anisotropy ratio of aluminum is small and the difference between the results from the Voigt and Reuss models is negligible. As seen in Fig. 1, the elastic moduli of aluminum significantly change with the temperature.

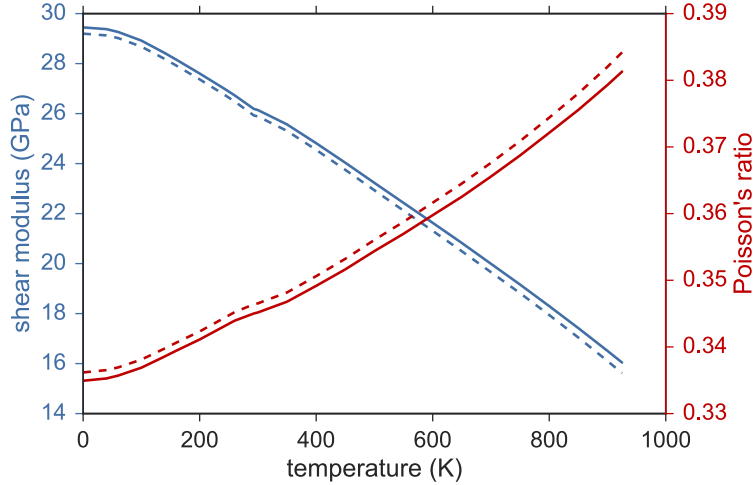

**Figure 1:** shear modulus, $\mu$, and Poisson's ratio, $\nu$ of pure polycrystalline aluminum versus temperature, $T$, calculated using the experimental results given in [17,18]. The results from the Voigt and the Reuss models are displayed by solid and dashed lines, respectively.

It has been clear that the drag coefficient is a function of temperature and pressure [21]. However, the way that $B$ varies with the temperature $T$ depends on the model we use [22]. In fcc crystals, which do not have considerable Peierls stress, the velocity of dislocations traveling through obstacles free zones is governed by phonon drag, except at very low temperatures where the interactions between dislocations and electrons become important [23,24]. Based on the fact that the dislocation is considered flexible or rigid, different terms contribute to phonon damping [25]. When a flexible dislocation line moves, it scatters phonons by fluttering. The fluttering mechanism gives rise to a force opposing the motion. The resulting drag coefficient is denoted by $B_{flut}$. When the dislocation is considered rigid, phonons become scattered by anharmonicities in the strain field. A moving dislocation experiences more phonon drag in the direction of motion than the opposite direction. The resulting drag is called the phonon wind, denoted by $B_{wind}$. The fluttering drag is assumed to vary with velocity, while $B_{wind}$ is regarded velocity independent. At low temperatures, $B_{wind}$ is proportional to $T^5$, at high temperatures both $B_{wind}$ and $B_{flut}$ are proportional to $T$ [25]. The temperature and velocity dependence of electron drag can be neglected [25]. Theoretically, for intermediate (say, $\Theta_D/3$, where $\Theta_D$ is the Debye temperature) to high temperatures, the total drag coefficient varies almost linearly with $T/\Theta_D$. Note that $\Theta_D$ =393 ± 1 K at 293 K, varying with the temperature to $\Theta_D$ = 362 ± 9 K at 559 K [26].

To find the variation of the drag coefficient of a dislocation in aluminum with the temperature, a nominally straight dislocation (of either edge or screw character) was considered under a constant external shear stress at a certain temperature. The edges of the computational box were along the $[1\bar{1}0]$, $[111]$ and $[11\bar{2}]$ crystal directions. For edge dislocations, periodic boundary conditions were applied along the dislocation line and the direction of the dislocation motion. For screw dislocations, periodic boundary conditions and free surfaces were applied along the dislocation line and the glide plane, respectively.

After relaxation at zero stress, dislocations dissociate into two partial dislocations. A constant shear stress was applied on the upper and lower (111) planes. To keep these surfaces flat, the

*y*-coordinates of some atomic planes adjacent to these surfaces were fixed. The embedded atom method (EAM) developed by Mishin et al. [27] was used in this study. This potential has been extensively used in many MD studies of aluminum [28–30] for it describes aluminum elastic properties and dislocations in aluminum well [29,31]. All calculations were carried out using the parallel MD code LAMMPS, developed in Sandia National Lab.

The potential we used results in almost zero Peierls stress for edge dislocations at $T \geq 1$ K [29]. The resulted Peierls stress for screw dislocations is ~ 30 MPa at $T = 1$ K [29], but it diminishes quickly at higher temperatures. The experimental results of the Peierls stress in fcc materials is typically small and hence is ignored in the calculations.

The values of the drag coefficient obtained from MD simulations for a screw and for an edge dislocation are plotted as a function of the absolute temperature, $T$, in Fig. 2. For example, at 300 K, the drag coefficient for the edge dislocation is $1.6 \times 10^{-5}$ Pa s, and that for the screw dislocation is $2.9 \times 10^{-5}$ Pa s. The drag coefficient for $T$<200 K and $T$>600 K varies approximately linearly with the temperature. The drag coefficient also monotonically increases with the temperature between 200 K and 600 K; $B'(T) = \partial B/\partial T$ reaches its maximum at $\approx 400$ K. Figure 2 shows a similar trend that was observed in the curve of $B$ versus $T$ for pure copper, compiled based on experimental results [32].

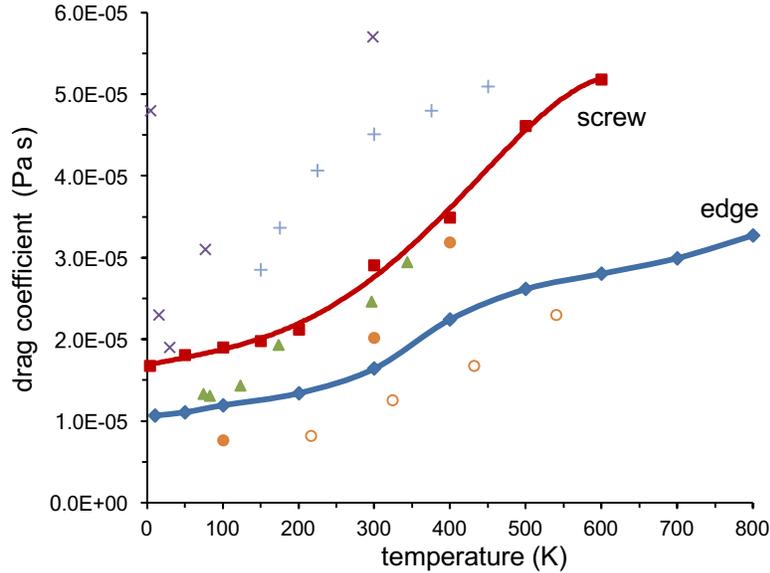

**Figure 2**: Drag coefficient for screw and edge dislocations in pure aluminum versus temperature *T*. MD simulation results of the present work are connected with solid lines. Experimental results by Gorman et al. (▲)[33,34] and Parameswaran et al. (✕) [35] are shown for comparison. MD results for edge dislocations (○) and for screw dislocations (●) by Olmsted et al. [22], and for edge dislocations (+) by Groh et al. [36] are also depicted. (with permissions from Elsevier and the American Institute of Physics)

The values of *B* are in good agreement with the experimental values given in Ref. [33,34]. The results of former MD simulations by Olmsted et al. [22] and Groh et al. [36] are also shown in Fig. 2. Compared with the previous MD studies, this work investigates the values of drag coefficients over a larger range of temperature, and has a better agreement with the values given in Ref. [33,34]. It, furthermore, studies the drag coefficients both for edge and screw dislocations, while the study by Groh et al. [36] is limited to only edge dislocations.

To model dislocation nucleation from Frank-Read (FR) sources in a 2d framework, we need to know (1) the minimum applied shear stress required to form a dislocation loop, $\tau_{\text{nuc}}$, and (2) the time it takes for the first dislocation loop to form, $t_{\text{nuc}}$. To characterize these parameters, a pure edge dislocation segment of length $L$ was considered under an external shear stress using a modified version of the 3d DDD code, MODEL. MODEL (Mechanics of defect evolution library) was originally developed at the University of California, Los Angeles, and was recently made open-source. For more details of this simulator, we refer readers to the paper by Po and Ghoniem [37]. We employed the nonsingular continuum theory of dislocations developed by Cai et al [38]. We also assumed that the drag coefficient of a mixed dislocation follows the following relationship [39]

$$\mathcal{B}(\boldsymbol{\xi}) = (B_s \cos^2 \theta + B_e \sin^2 \theta)(\boldsymbol{I} - \boldsymbol{\xi} \otimes \boldsymbol{\xi}) \tag{2}$$

where $\boldsymbol{\xi}$ is the dislocation line direction, $\boldsymbol{I}$ is the identity tensor, $\theta$ is the angle between $\boldsymbol{\xi}$ and the Burgers vector $\boldsymbol{b}$, and $B_s$ and $B_e$ are the drag coefficients for screw and edge dislocations, respectively.

Using 3d DDD and the aforementioned assumptions, we measured the nucleation stress $\tau_{\text{nuc}}$ for several lengths of the dislocation segment, $L$, at different temperatures. Foreman [40] suggested that the relationship between $\tau_{\text{nuc}}$ and $L$ takes the following form:

$$\tau_{\text{nuc}} = \frac{\mu b}{2\pi(1-\nu)L}\left[A \ln\left(\frac{L}{b}\right) + B\right] \tag{3}$$

where $A$ and $B$ are two constants. To verify Eq. (3) for our data, the values of $\tau_{\text{nuc}}$ from simulations in terms of $\mu/(1-\nu)$ are plotted versus $L/b$ (Fig. 3(a)). The solid line in Fig. 3(a) represents $\tau_{\text{nuc}}$ from Eq. (3) when this equation was fitted to the values of $\tau_{\text{nuc}}$ at 300 K and obtained $A = 0.765$ and $B = -1.330$. The errors of this model in this figure, is less than $\approx$ 2--3 MPa at different temperatures. However, the nucleation stress starts to deviate from Eq. (3) when $L < 180\ b$. Because of the distribution we select for the length of FR sources, the probability of finding an FR source of length smaller than $180b$ is very small.

Nucleation time as a function of the reduced applied shear stress, $\tau/\tau_{nuc}$, has been plotted at different temperatures when the initial length of the source $L$ is $220b$ (Fig. 3b) and for different initial lengths of the source when the temperature is 300 K (Fig. 3c). When $1 \leq \tau/\tau_{nuc} \leq 1.05$, nucleation time drastically increases to a few hundreds of nanoseconds; this interval was not shown in Fig 3b and Fig. 3c. Nucleation time has been defined as the time interval to reach critical configuration by Benzerga [41] and Gómez-García et al. [42]. This time interval calculated by Benzerga [41] and Gómez-García et al. [42] underestimate and overestimate the nucleation time, respectively, compared to what we measured computationally. When studying the rate sensitivity of yield strength and work-hardening, it is important to take into account the realistic nucleation time [43].

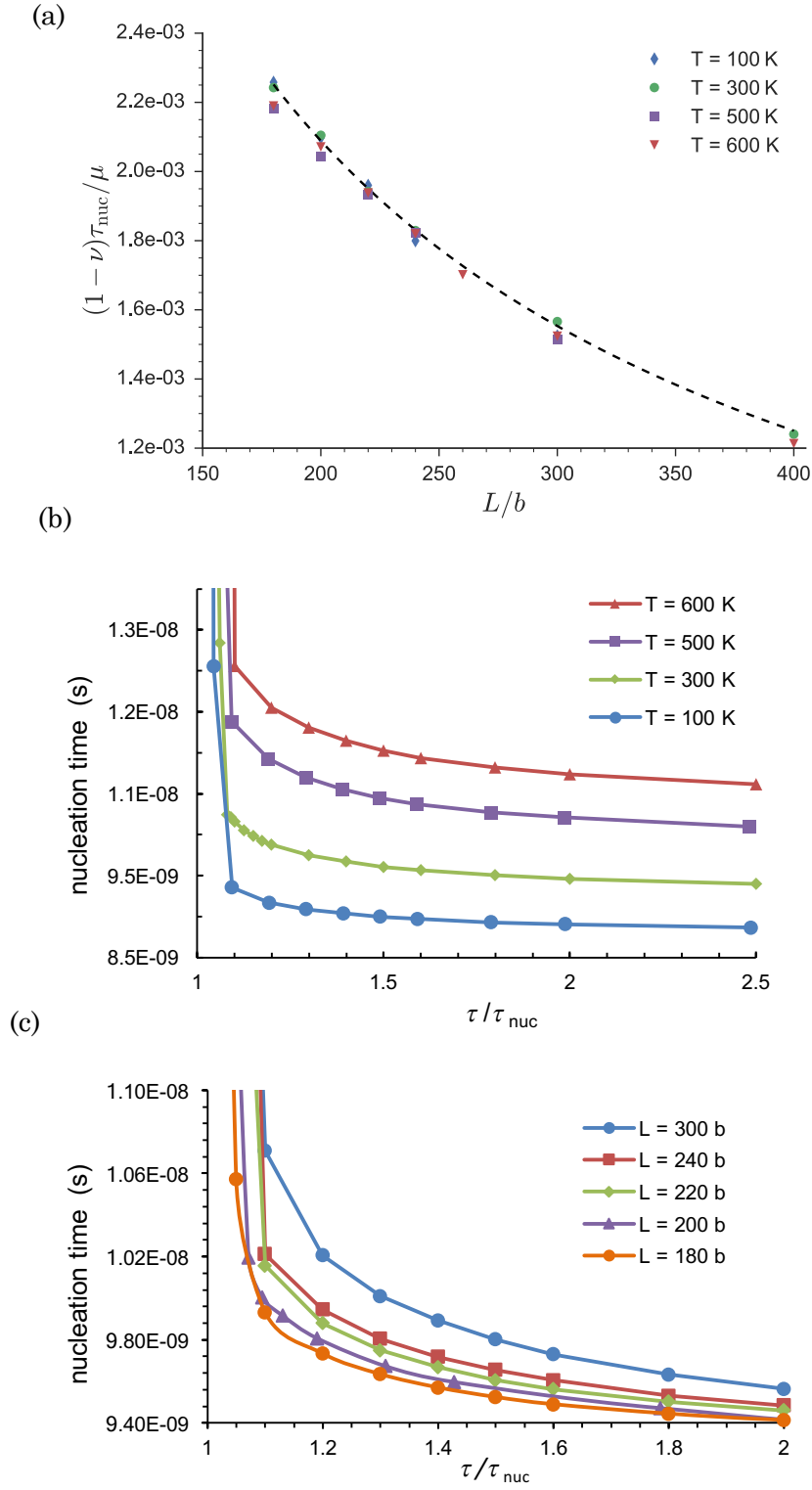

**Figure 3:** (a) critical nucleation stress versus the length of the Frank-Read source at different temperatures. Nucleation time for different values of applied stress $\tau/\tau_{\text{nuc}}$ (b) at different temperatures when $L = 220b$ (c) for different values of $L$ when $T=300$ K.

To investigate the effects of the parameters studied here on the yield stress $\sigma_y$, we considered an infinitely long aluminum thin film of thickness 0.75 μm under uniaxial tension in a 2d DDD simulation. Tension was applied by prescribing a constant displacement rate. The film was passivated on both sides, and grains to be columnar with the size of 1.0 μm. Periodic boundary condition was assumed along the *x*-axis; grains were randomly oriented. No internal obstacles were present to impede the dislocation motion. As shown before [39,44] and confirmed by our MD calculations, after a time that is smaller than the typical time increment in DDD calculations, dislocations reach a steady state velocity. Therefore, as long as the dislocation velocity is a fraction of the speed of sound, the inertia effects can be neglected and Eq. (1) reduces to $Bv = F(t)$. The new location of a dislocation was determined by the Euler-backward integration [45]. For more details about the simulation method, we refer interested readers to Ref. [4,10,11].

The lengths of FR sources were randomly chosen from a Gaussian distribution with the mean value of $220b$ and standard deviation of $20b$. The density of FR sources was taken to be 13 μm$^{-2}$. Films were initially dislocation free, but FR sources were randomly distributed in slip planes. Critical nucleation stress of each source, $\tau_{nuc}$, was calculated from Eq. (3). Testing for normality of the empirical distribution of $\tau_{nuc}$ was performed with the Kolmogorov-Smirnov and Shapiro-Wilk tests. These tests did not provide enough evidence against the null hypothesis that the empirical distribution of $\tau_{nuc}$ was normal. Also Q-Q plots showed the distribution to be close to the normal.

Curves of the *xx*-component of the stress field averaged over the film thickness versus the applied strain are depicted in Fig. 4. The slope of the elastic loading is given by the plain-strain modulus of the film, $E/(1-v^2)$. In this study, the plane-strain yield stress, $\sigma_y$, is defined at an offset plastic strain of 0.2%, and obtained from the intercept of the dashed lines in Fig. 4, which are parallel to the initial straight-line portions of the stress-strain curves, with the computed stress-strain curves. The values of the yield stress at different temperatures are shown in the inset of Fig. 4. Yield stress decreases by ≈13%, when the temperature increases from 100 K to 600 K; the plain strain modulus decreases by ≈18%.

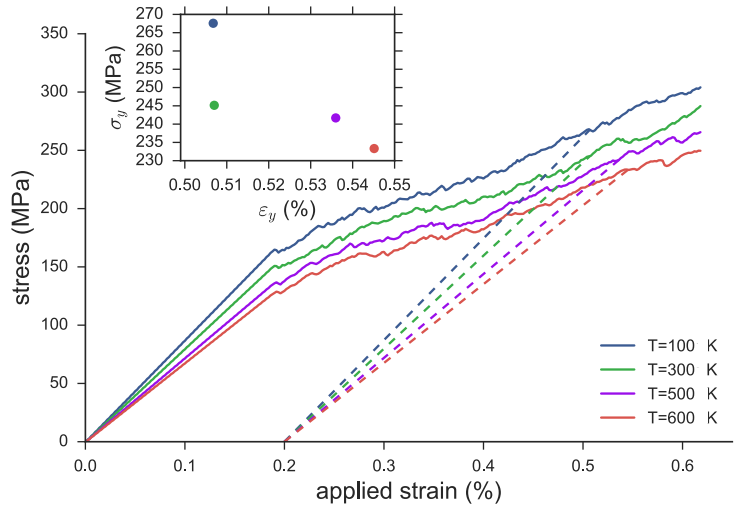

**Figure 4:** stress-strain curve for aluminum thin film of thickness 0.75 μm at different temperatures. The inset shows variation of the yield stress $\sigma_y$ versus strain $\varepsilon_y$ with the temperature.

The values of the source density, the mean and standard deviation of the FR source length were the typical numbers found in the literature. However, the onset of plastic deformation in the stress-strain curve and the yield stress are mostly determined by the distribution of the nucleation stress and the hardening rate by the FR source density [46]. Similar outcomes have been reached through experiments as well. Experiments have indicated that the onset of macroscopic plastic deformation is mainly controlled by dislocation nucleation and multiplication; *in situ* transmission electron microscopy (TEM) studies have showed that prior to the yield point, many dislocations already exist in the sample and the yield event is related to the rearrangement of these dislocations [47]. Therefore, the distribution of the nucleation stress and the density of FR sources can be set by fitting the experimental stress-strain curves at a certain temperature. Then, from the variations we investigated in this paper, the stress-strain response at different temperatures can be estimated.

Dislocation climb rarely occurs before the yield point even when $T = 600$ K; thus, its effect is negligible here. Nevertheless, climb is an important recovery mechanism at larger strains (see, e.g., [4,10]). Temperature dependence of the vacancy migration energy and that of the pre-exponential coefficient of diffusivity in aluminum are very small [48]. Fluss et al. [49] determined the vacancy formation enthalpy as 0.66±0.02 eV for temperatures below 645 K. With this information, dislocation climb at different temperatures can be readily implemented.

In this letter, we focused on pure aluminum as an fcc material, but we should note that different approaches may be needed for many other materials. In general, dislocation velocity is affected by extrinsic factors, such as impurities which often exist in reality, and by intrinsic factors, such as lattice resistance to dislocation motion [39]. In materials where the Peierls stress is high, dislocations move by kink-pair mechanisms [24]. In such cases, dislocation velocity is basically determined by the thermally activated overcoming of the lattice resistance, which may lead to a nonlinear mobility law [1].

In summary, we have investigated the effects of temperature on the elastic constants, nucleation stress and nucleation time. To this end, we have used the existing experimental results, MD and 3d DDD simulations. We have shown if we take these variations into account, the yield stress can be considerably influenced.

The author would like to thank Prof. Wei Cai of Stanford University for an insightful discussion and Dr. Giacomo Po of UCLA for his help with MODEL.